\documentclass{article}

\usepackage{microtype}
\usepackage{graphicx}
\usepackage{subcaption}
\usepackage{booktabs} % for professional tables

\usepackage{hyperref}

\usepackage[preprint]{icml2026}

\usepackage{amsmath}
\usepackage{amssymb}
\usepackage{mathtools}
\usepackage{amsthm}

\usepackage{booktabs}
\usepackage{multirow}
\usepackage{tcolorbox}       % 核心包：彩色盒子

\usepackage[capitalize,noabbrev]{cleveref}

\theoremstyle{plain}

\theoremstyle{definition}

\theoremstyle{remark}

\usepackage[textsize=tiny]{todonotes}
\tcbuselibrary{breakable}%prompt分页
\usepackage{enumitem}
\icmltitlerunning{EmoOmni: Bridging Emotional Understanding and Expression in Omni-Modal LLMs}

\begin{document}

\twocolumn[
  \icmltitle{EmoOmni: Bridging Emotional Understanding and Expression \\
  in Omni-Modal LLMs}



  \icmlsetsymbol{equal}{*}

  \begin{icmlauthorlist}
    \icmlauthor{Wenjie Tian}{equal,yyy}
    \icmlauthor{Zhixian Zhao}{equal,yyy}
    \icmlauthor{Jingbin Hu}{yyy}
    \icmlauthor{Huakang Chen}{yyy}
    \icmlauthor{Haohe Liu}{comp}
    \icmlauthor{Binshen Mu}{yyy}
    % \icmlauthor{Xiaohai Tian}{comp}
    % \icmlauthor{Jun Zhang}{sch}
    % \icmlauthor{Lu Lu}{yyy,comp}
    % \icmlauthor{Yuxuan Wang}{yyy,comp}
    \icmlauthor{Lei Xie}{yyy}
  \end{icmlauthorlist}

  \icmlaffiliation{yyy}{Northwestern Polytechnical University, Xi'an, China}
  \icmlaffiliation{comp}{University of Surrey, Guildford, United Kingdom}

  \icmlcorrespondingauthor{Wenjie Tian}{twj@mail.nwpu.edu.cn}
  \icmlcorrespondingauthor{Lei Xie}{lxie@nwpu.edu.cn}

  \icmlkeywords{Machine Learning, ICML}

  \vskip 0.3in
]



\printAffiliationsAndNotice{\icmlEqualContribution}

\begin{abstract}
The evolution of Omni-Modal Large Language Models~(Omni-LLMs) has revolutionized human–computer interaction, enabling unified audio-visual perception and speech response. 
However, existing Omni-LLMs struggle with complex real-world scenarios, often leading to superficial understanding and contextually mismatched emotional responses.
This issue is further intensified by Omni-LLM's Thinker-Talker architectures, which are implicitly connected through hidden states, leading to the loss of emotional details.
In this work, we present EmoOmni, a unified framework for accurate understanding and expression in multimodal emotional dialogue. 
At its core, we introduce the emotional Chain-of-Thought~(E-CoT), which enforces a reasoning from fine-grained multimodal perception to textual response. 
Moreover, we explicitly treat E-CoT as high-level emotional instructions that guide the talker, enabling accurate emotional expression.
Complementing the model, we construct EmoOmniPipe to obtain the real-world annotated dialogue data and establish a benchmark, EmoOmniEval, to facilitate systematic assessment of multimodal emotional dialogue task.
Experiments show that EmoOmni-7B achieves comparable performance with Qwen3Omni-30B-A3B-Thinking under the same talker. 
% Projects are open-sourced in~\url{https://anonymous.4open.science/r/EmoOmni-7525/}.

\end{abstract}

\section{Introduction}
\label{intro}

\begin{figure}[t]
  \centering
  \includegraphics[width=1\linewidth]{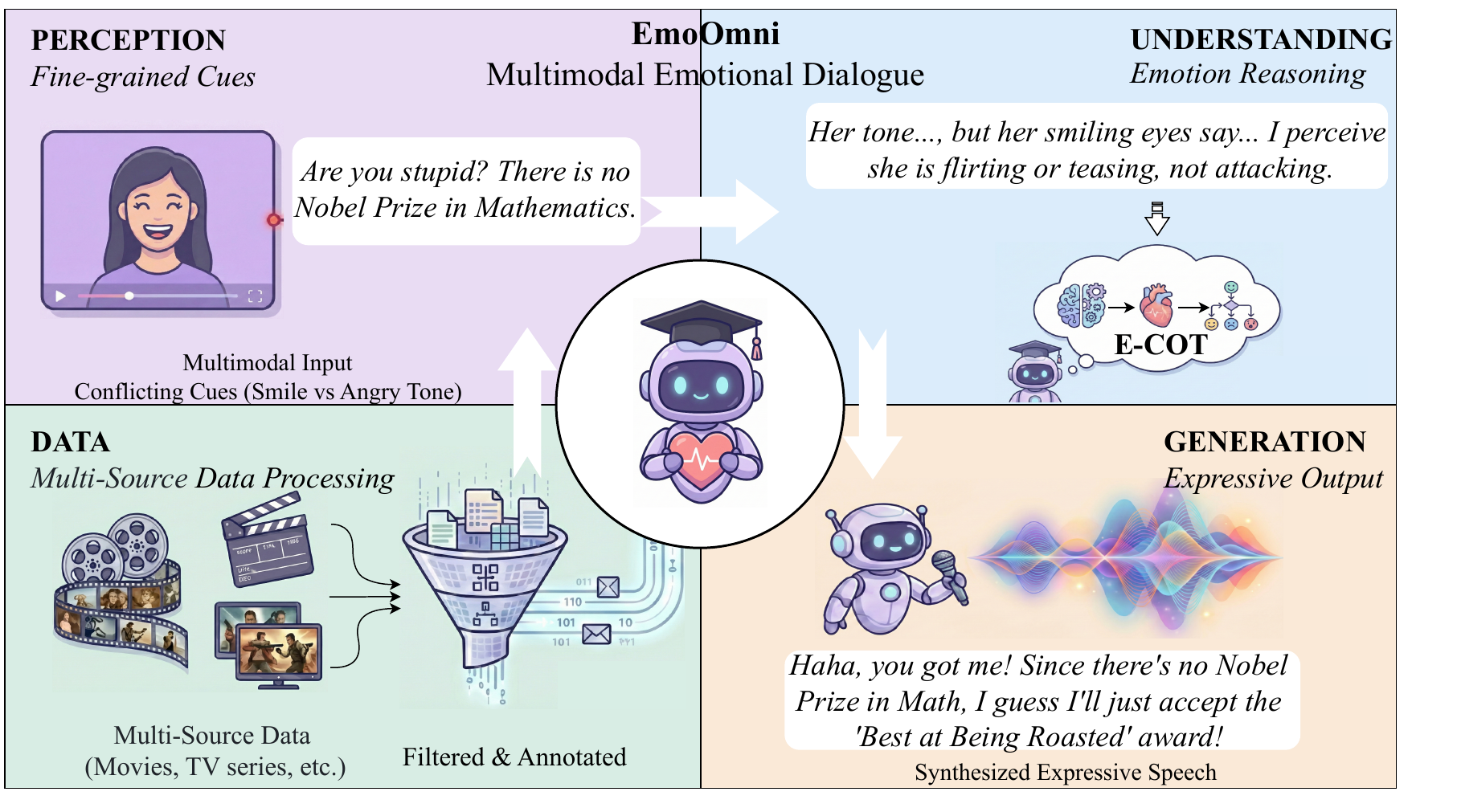} 
  \caption{The overall framework of EmoOmni. The system mimics human affective cognition through a Perception-Reasoning-Expression causal chain. }
  \label{overall}
\end{figure}
In human–computer interaction~(HCI), emotional intelligence plays a critical role in enabling natural and effective dialogue.
With the emergence of Omni-Modal Large Language
Models (Omni-LLMs)~\cite{qwen3omni,Interns1,gemini2.5,mingomni,minicpm,gpt4o}, HCI has evolved from text-only communication~\cite{qwen3} toward multimodal dialogue involving fused audio-visual perception and speech response.
Under this paradigm shift, achieving emotionally intelligent interaction requires models to jointly perceive audiovisual cues rather than relying solely on text.
They need to accurately infer the user’s underlying intent and emotional state, and generate speech responses that are emotionally appropriate to the interaction context. 
However, achieving such remains challenging due to limitations spanning model capability, data resources, and evaluation methodologies. 

% 困难1
% 真实场景复杂，甚至会有emotion 冲突的时候，模型很难处理。（但是重点在真实对话中）
In real-world conversational interactions, existing Omni-LLMs fail to handle these realistic scenarios because the audio and visual cues are often complex, implicit, or even conflicting across modalities.
For example, a speaker may express a cheerful tone while displaying a frowning facial expression, as illustrated in Figure~\ref{overall}. 
As a result, they often rely on superficial understanding and produce incorrect intent inference.
This misinterpretation propagates to the generation stage, leading to responses that are inappropriate or socially misaligned in real conversational contexts.
More critically, even when the underlying intent and emotions are accurately inferred, the speech ultimately generated by the model may still fail to reflect the desired emotional state. 
This is because current Thinker–Talker architectures in Omni-LLMs solely depend on implicit emotional control through hidden states.
To maintain semantic correctness, this design inevitably leads to the loss of emotional details,
where emotional intent is diluted or lost during transmission.
As a result, expressive speech synthesis is compromised, the final speech output may be semantically appropriate but emotionally misaligned, such as conveying reassurance without warmth.

% 困难3
% 数据和benchamrk.
Fundamentally, the scarcity of real-world, well-annotated multimodal dialogue data poses a major bottleneck. 
Although raw digital media is abundant, no widely adopted pipeline currently exists. Furthermore, existing dialogue datasets~\cite{multidialog, meld} either lack labels or are limited to coarse tags~\cite{labeldata_dialogue1} such as ``happy", failing to provide detailed annotations that help models capture human interactive behaviors.
Finally, most existing benchmarks~\cite{MMLA, emobench-m} focus on task correctness or basic emotion recognition accuracy, overlooking whether the generated responses demonstrate emotional intelligence within the interaction context.
This gap limits meaningful evaluation.

%总结
To address these challenges, we introduce EmoOmni, a higher emotional intelligence framework in Multimodal Emotional Dialogue (MED).
Unlike traditional Omni-LLMs, EmoOmni mimics human through a Perception–Reasoning–Expression causal chain, explicitly disentangling emotional understanding, strategic decision-making, and acoustic expression.
Specifically, at the perception stage, EmoOmni enhances fine-grained multimodal perception to extract subtle acoustic and visual signals. 
Building on better perception, we introduce the Emotional Chain-of-Thought (E-CoT) mechanism to enable deliberate reasoning that it first infers the user’s intent and emotion.
Then it generates an appropriate dialogue strategy under the conversational context to achieve semantic alignment.
This strategic reasoning explicitly serves as an  emotional instruction for  EmoOmni-Talker, ensuring that the final expression is semantically and emotionally aligned with the context.
% is not only semantically coherent but also acoustically aligned with the context. 
Complementing the architecture, we adopt a two-stage training strategy that progressively cultivates perception and reasoning abilities.
To address data scarcity and support learning across the entire Perception–Reasoning–Expression chain, we construct a data pipeline, EmoOmniPipe, that extracts and annotates emotionally rich dialogues from movies and TV series.
Finally, to systematically evaluate emotional intelligence in multimodal interaction, we establish EmoOmniEval, a comprehensive benchmark covering perception, reasoning, and expressive generation.
Remarkably, our 7B parameter model achieves parity with 30B scale models, demonstrating that E-CoT and real-world data can effectively compensate for parameter scale in affective computing.
Our main contributions are summarized as follows:
\begin{itemize}
\item  \textbf{Framework:} We propose EmoOmni, a novel Omni-LLM framework that explicitly models MED as a \emph{Perception–Reasoning–Expression} causal chain, enabling more accurate understanding and expression.
\item \textbf{Methodology:} 
We introduce the E-CoT not only as an explicit emotional reasoning process that bridges perception and expression but also as explicit instructions to guide EmoOmni-Talker to preserve emotional details.
They both ensure the final response is semantically and emotionally aligned with the context.
\item  \textbf{Data and Benchmark:} We construct a MED data pipeline and a benchmark. The EmoOmniPipe data pipeline provides comprehensive processing and annotation to support training.  The benchmark, EmoOmniEval, systematically evaluates model's capability of emotional understanding and expression.
\item \textbf{Performance:} Experiments show that despite using only 7B parameters, EmoOmni achieves performance comparable to 30B-scale Omni-LLMs under the same talker, demonstrating that explicit emotional reasoning and instruction-guided expression can effectively compensate for model scale in MED.

\end{itemize}

\section{Related Work}
\subsection{Omni-LLMs}
% Omni-Modality and Human-Computer Interaction
Current Omni-LLMs~\cite{gemini2.5,Interns1,qwen3omni, mingomni} represent unified multimodal frameworks capable of simultaneously processing text, audio, and video inputs while engaging in dialogue through speech responses. 
Architecturally, these systems generally consist of a Thinker module responsible for multimodal comprehension, followed by a talker module that converts textual responses into speech. 
However, these models are primarily optimized for general understanding and response fluency, rather than fine-grained affective reasoning and expressive alignment.
Some works~\cite{mllm_1,mllm_2} have taken emotion modeling into account, yet they are limited to using a single emotion tag.
Consequently, they struggle to disentangle conflicting multimodal cues and fail to maintain deep affective alignment, rendering them insufficient for scenarios requiring precise intent understanding and strategic empathy.

\subsection{Multimodal Emotion Comprehension}

Research in this domain has shifted from static Multimodal Emotion Recognition~\cite{merbench} to generative Multimodal Emotion Reasoning~\cite{AffectGPT}. Early works~\cite{unimse, mmgcn} focus on feature fusion for discrete label classification, which limits their applicability in open-world scenarios. The advent of LLMs introduces generative reasoning paradigms: for example, Emotion-LLaMA~\cite{emotion-llama, merllm, merllm2} pioneered emotion-specific instruction tuning. To address the challenge of uni-modal dominance in conflicting scenarios, recent works have proposed sophisticated fusion mechanisms, such as attention reallocation in MOSEAR~\cite{mosear} and gate-fusion in HumanOmni~\cite{humanomni}. Most notably, SABER~\cite{saber} utilizes structured evidence decomposition to achieve robust reasoning. However, a critical gap remains as most existing works focus exclusively on the analysis phase. They treat emotion understanding as an isolated task, failing to integrate the perception and expression.

\subsection{Emotional Speech Generation}

Expressive speech synthesis has progressed from label-based methods to natural description control.
Although label-based approaches~\cite{stepaudio2,indextts2} enable attribute control via tags, they are constrained by a predefined and limited set of style categories.
While prompt-based systems~\cite{instructTTS,VoiceSculptor} allow textual control over prosody, they are often restricted to predefined style descriptions or broad emotional categories. In the context of Omni-models, the talker module typically receives implicit, high-dimensional control signals from the Thinker. 
These models struggle to achieve acoustic–semantic alignment, often generating speech that is semantically correct but acoustically generic, failing to reflect the dynamic emotional strategies, such as soothing and emphasizing. These are important for natural conversation.
% 仍然有效，并且在情感上表现不强。

\begin{figure*}[t!]
  \centering
  \includegraphics[width=1\linewidth]{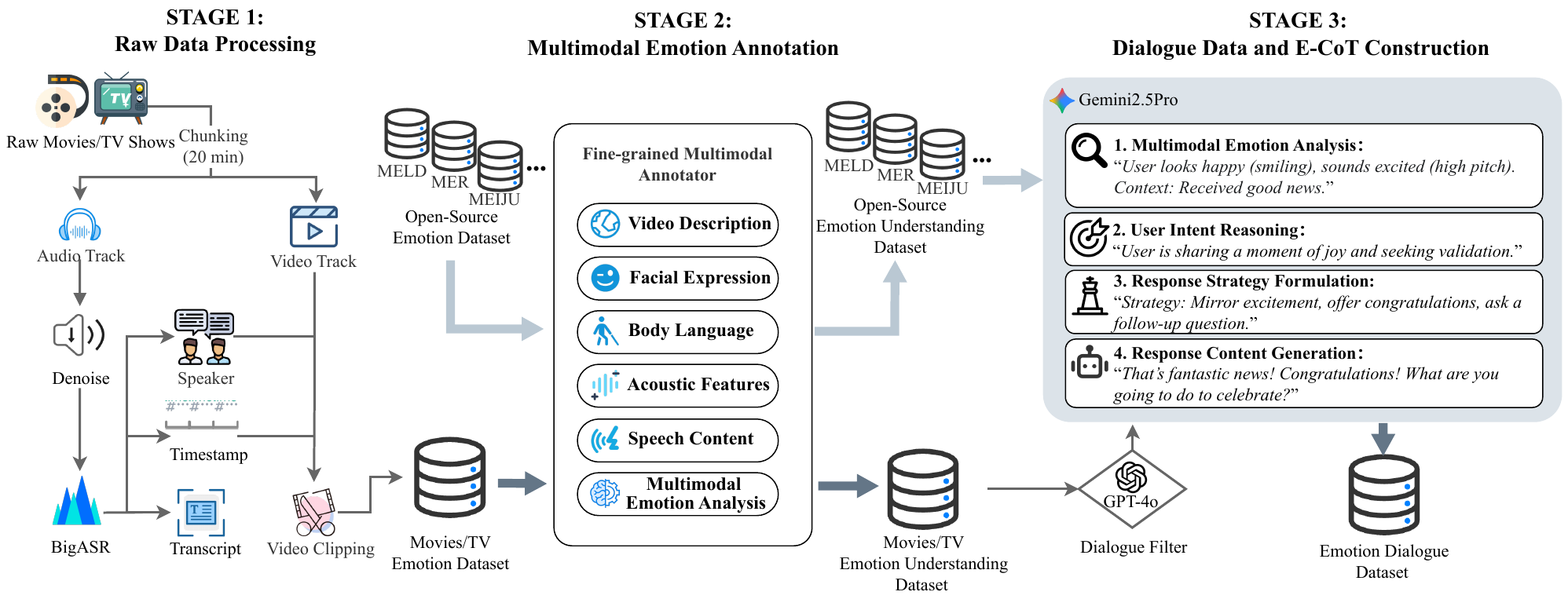}  % 使用单栏宽度的80%
  \caption{Overview of the EmoOmniPipe.
   The data pipeline consists of three main stages: (1) processing of raw data, (2) multimodal annotation, and (3) construction of E-CoT and dialogue data.
  }
  \label{data}
\end{figure*}

% nmos, n-mos, i-mos(可懂度), 
\section{EmoOmniPipe and EmoOmniEval}
% data construction and benchmark
To enable a holistic, fine-grained, and reproducible evaluation of MED systems, we introduce a real-world dialogue data processing pipeline, EmoOmniPipe, and a multidimensional assessment benchmark, EmoOmniEval.

\subsection{Real-World Data Processing Pipeline}
To capture the rich emotional dynamics inherent in genuine human interaction, we select movies and TVs as our primary data sources. 
To address unstructured data, we designed a data cleaning and processing pipeline to ensure the construction of a high-quality corpus as shown in stage 1 of Figure~\ref{data}.
We first segment raw videos into 20-minute chunks and isolate audio and video tracks.
Subsequently, we employ audio with MelBandRoformer~\cite{melbandroformer} for denoising, thereby significantly improving acoustic quality.
Upon obtaining the clean audio, we leverage the Volcengine BigASR API~\footnote{\url{https://www.volcengine.com/docs/6561/1354869?lang=zh}} to extract precise segmentation timestamps, speaker diarization labels, and  transcripts. 
To maintain conversational continuity, we merge consecutive segments from the same speaker and perform precise video clipping based on their timestamps.  

During stage 2, we leverage state-of-the-art (SOTA) open-source models~\cite{saber} to annotate the processed items. The models generate fine-grained multimodal captions including six dimensions. 
After annotating, we split continuous context into appropriate dialogue segments using GPT4o~\cite{gpt4o}, filtering the dialogues lacking sufficient contextual information or containing negative content. The detailed filtering criteria and corresponding prompts are provided in the Appendix~\ref{app_sec_filter}.
This systematic pipeline results in a large-scale, high-fidelity dataset of single-turn dialogues in the real world.

\subsection{E-CoT Construction}
\label{ecot_method}
After finishing the process and annotation of dialogue data, we construct a structured E-CoT comprising four key components: multimodal emotion analysis, user intent recognition, response strategy planning, and response content.
Specifically, as shown in stage 3 of Figure~\ref{data}, by feeding the Gemini2.5Pro~\cite{gemini2.5} the contextual history and the annotated multimodal emotional analysis, we prompt it to deductively infer the thinking process, leading to the target response.
To further expand the scale and diversity of our training data, we apply this method to existing open-source multimodal emotion understanding~(MEU) datasets. By providing groundtruth video and emotional analysis, we guide the model to generate E-CoT with generated response content.

% \subsection{Multidimensional Evaluation Benchmarks}
\subsection{EmoOmniEval}

We utilize MELD-test (English) and ch-sims-v2 (Chinese) as test sets for MED tasks. 
Both of them undergo the complete processing pipeline to generate the E-CoT. 
Given that open-ended dialogue tasks are difficult to quantify using traditional metrics, we employ LLM-as-a-Judge, leveraging Gemini2.5Pro to implement a decoupled multidimensional assessment system. 
While directly evaluating the correlation between the input video and the final speech output offers an end-to-end perspective, relying solely on this dimension fails to pinpoint specific weaknesses in the model’s internal processing.
To enable more fine-grained analysis, we introduce additional evaluations based on intermediate textual responses and instruction-driven speech synthesis.
Specifically, our evaluation framework consists of three complementary settings:
(1) end-to-end evaluation from input video to generated speech,
(2) evaluation of textual responses generated from input video, and
(3) evaluation of speech synthesis conditioned on explicit instructions.
For clarity, we refer to these settings as \textbf{\emph{Video-to-Speech~(VS), Video-to-Text Response~(VT), and Instruction Following~(IF)}}, respectively.
All metric scores are graded 0, 1, or 2. See the Appendix~\ref{benchmarkprompt} for details on the prompts.

\textbf{VS Evaluation:} This dimension assesses the overall performance of the system. We evaluate Response Content Relevance and Logic~(VS-RC), which determines whether the speech output semantically aligns with the video input and maintains logical coherence. Additionally, we assess Response Emotional Strategy~(VS-RES), which measures whether acoustic features of the generated speech are an appropriate reaction to the user's emotional state depicted in the video.

\textbf{VT Evaluation:} To isolate the reasoning capabilities, we evaluate text generation quality. This includes Emotion Analysis~(VT-EA) accuracy, which judges the completeness and correctness of the multimodal emotion analysis. We also evaluate the textual version of Response Emotional Strategy~(VT-RES) and Response Content Relevance and Logic~(VT-RC) to ensure correct reasoning.

\textbf{IF Evaluation:} Inspired by Instruct-TTS-eval, IF measures how accurately the speech generation system adheres to the provided emotional and acoustic instructions, assessing the capability of the talker module.

\begin{figure*}[t!]
  \centering
  \includegraphics[width=1\linewidth]{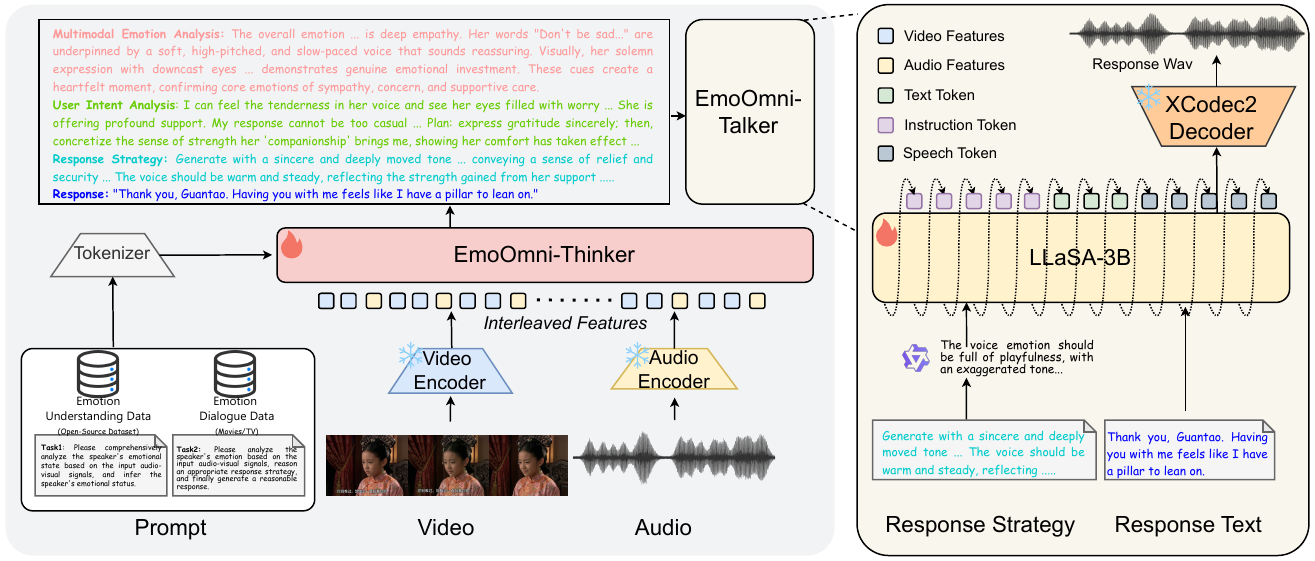}  % 使用单栏宽度的80%
  \caption{Overview of EmoOmni.
  The left part illustrates the overall pipeline of our multimodal emotional dialogue system. 
  % During multitask training, we leverage generated data covering two tasks: multimodal emotion understanding and multimodal emotional dialogue. 
  Given audio–visual inputs, the Thinker module performs high-level reasoning and produces the emotional Chain-of-Thought, which consists of four components.
  The generated textual response is then fed into the talker module for expressive speech synthesis.
  The right part shows the architecture of the talker, which is built upon an autoregressive TTS model. The response strategy is further processed by a lightweight language model to generate controllable emotion instructions, enabling the talker to synthesize  expressive speech.
  }
  \label{model}
\end{figure*}
\section{EmoOmni Framework}

\subsection{Overall Architecture}

EmoOmni aims to address the challenge of emotionally intelligent multimodal dialogue by decomposing the generation process into distinct perception, reasoning and expression stages. 
Given multimodal inputs consisting of video $V$ and audio signals $A$, the framework aims to generate an emotionally appropriate speech response $Y$ that aligns both semantically and affectively with the user's state. This process can be formulated as:
\begin{equation}
\hat{Y} = \text{EmoOmni}(V, A).
\end{equation}
To achieve this goal, EmoOmni is designed to explicitly mirror the human Perception–Reasoning–Expression causal chain.
As illustrated in Figure~\ref{model}, the framework consists of two yet tightly coordinated modules:
(i) \textbf{EmoOmni-Thinker} that performs multimodal perception and reasoning, and
(ii) \textbf{EmoOmni-Talker} that converts the reasoned textual response into emotionally expressive speech.
The thinker first transforms multimodal inputs into a structured reasoning trajectory, while the talker converts the resulting strategy into controllable acoustic outputs.
This explicit separation allows EmoOmni to disentangle \emph{what to say} from \emph{how to say it}, which is essential for achieving reliable intent and affect alignment in complex real-world scenarios.

\subsection{Multimodal Encoding and Fusion}
Based on Qwen2.5-Omni's multimodal encoders, EmoOmni first encodes the raw media $V$ and $A$ into continuous feature spaces, which are then interleaved temporally to form a fused multimodal stream $\mathcal{M}$. 
This unified representation $\mathcal{M}$ serves as the perceptual foundation for subsequent reasoning.
This method provides a shared semantic space where cross-modal emotional cues can be jointly interpreted.

\subsection{E-CoT mechanism}
Rather than treating dialogue generation as a black-box mapping from multimodal inputs to responses, EmoOmni explicitly models the reasoning process that connects perception to expression. 
Given multimodal inputs $\mathcal{M}$, E-CoT models emotional dialogue generation as a structured reasoning trajectory consisting of three latent cognitive stages: \textbf{\emph{Perception}, \emph{Intention Analysis}, and \emph{Response Strategy}}. Each stage incrementally refines the model's understanding before producing the final response. 
Crucially, this process enforces a causal dependency chain that each stage conditions the next in an autoregressive manner, from perception to expression.
% 上面这句话很重要，因为是ar的，所以是causal的。

\noindent\textbf{Multimodal Emotion Perception.}
The reasoning chain begins with fine-grained emotional perception. Conditioned on multimodal features, the model describes observable emotional cues from audio and video, such as vocal tension, facial expressions, or behavioral inconsistencies. 
This perceptual representation grounds subsequent reasoning in factual multimodal evidence, reducing hallucinated interpretations.
Formally, this stage corresponds to the conditional probability $P(z_p|\mathcal{M})$, where $z_p$ denotes the perceived emotional state derived from inputs $\mathcal{M}$.

\noindent\textbf{Intention Analysis.}
Based on perceived emotional cues $z_p$, the model infers the user's underlying intention and mental state. 
This step bridges surface-level observations and latent motivations, allowing the model to resolve complex phenomena such as sarcasm, emotional masking, or conflicting multimodal signals. The inference process is captured by
$
P(z_a|z_p)
$
, where $z_a$ represents the analytically inferred intention conditioned on perception $z_p$.

\noindent\textbf{Response Strategy Planning.}
Given the inferred intention $z_a$, the model explicitly plans a high-level response strategy that determines how the system should react emotionally and pragmatically. 
This strategy functions as an abstract control signal, decoupling decision-making from surface realization. The planning stage is modeled as:
$
P(z_s|z_a)
$
, where $z_s$ denotes the selected response strategy based on intention analysis $z_a$.

\noindent\textbf{Textual Response Generation.}
Finally, the system generates the textual response $z_t$ conditioned on the planned strategies $z_p, z_a, z_s$.
Complete E-CoT $Z = \{z_p, z_a, z_s, z_t\}$ ensures that the model correctly infers the user's state and plans the appropriate response. 
Denote the parameter of thinker as $\theta^{\text{Thinker}}$, and the causal generation chain is as follows: 
\begin{equation}
P(Z|\mathcal{M}, \theta^{\text{Thinker}}) = P(z_p|\mathcal{M}) P(z_a|z_p) P(z_s|z_a) P(z_t|z_s).
\end{equation}

\subsection{Two-Stage Training Strategy}

The efficacy of the proposed causal chain relies heavily on the accuracy of the initial perception node $z_p$. 
In the causal chain, a deviation in the initial perception inevitably leads to cascading errors in subsequent reasoning and generation steps. 
To mitigate this, we design a progressive two-stage training paradigm that optimizes perceptual grounding before reasoning.

\noindent\textbf{Stage 1: Perceptual Grounding.} 
This stage focuses exclusively on optimizing the perception term $P(z_p|\mathcal{M})$. 
By leveraging our constructed MEU dataset, we finetune the Thinker module to align its feature space with fine-grained emotional concepts. 
 We minimize the negative log-likelihood (NLL) of the perception variable $z_p$:
\begin{equation}
    \mathcal{L}_{\text{Stage1}} = -\mathbb{E}_{(\mathcal{M}, z_p) \sim \mathcal{D}_{\text{MEU}}} [\log P(z_p | \mathcal{M}); \theta^{\text{Thinker}}],
\end{equation}
where $\mathcal{D}_{\text{emo}}$ is our constructed MEU dataset.
This step maximizes the mutual information between multimodal inputs and emotional descriptions, establishing a solid foundation for the subsequent reasoning chain.

\noindent\textbf{Stage 2: Joint Reasoning Tuning.} 
Building upon the calibrated perception, the second stage activates the full causal chain. 
We utilize MED and MEU tasks to jointly optimize the entire sequence from perception $z_p$ to generation $\hat{Y}$.
Since the perception module is already stabilized, the model can focus on learning the complex causal dependencies $P(z_a|z_p)$ and $P(z_s|z_a)$. 
The objective function explicitly decomposes into four causal components:
\begin{equation}
\begin{aligned}
    \mathcal{L}_{\text{Stage2}} = - \mathbb{E}_{(\mathcal{M}, Z) \sim \mathcal{D}_{\text{MEU+MED}}} \Bigg[ 
     \underbrace{\log P(z_p | \mathcal{M})}_{\text{Emotion Analysis}} \\
    +  \underbrace{\log P(z_a | z_p)}_{\text{Intent Analysis}} 
    + \underbrace{\log P(z_s | z_a)}_{\text{Response Strategy}} +  
    \underbrace{\log P(z_t | z_s)}_{\text{Response Generation}} 
    \Bigg].
\end{aligned}
\label{eq:stage2_loss}
\end{equation}
This curriculum learning approach successfully transitions the thinker from perception to reasoning and effectively ensures that the generation is genuinely driven by the deep multimodal understanding established in the first stage.

\subsection{Instruction-Guided Speech Generation}
The final stage addresses acoustic-semantic alignment by modeling $P(\hat{Y}|z_t, z_s)$. We develop EmoOmni-Talker, an instruction-guided Text-to-Speech (TTS) system built on VoiceSculptor~\cite{VoiceSculptor}.
It transforms the thinker's high-level strategy $z_s$ and textual response $z_t$ into expressive waveforms $\hat{A}$. Specifically, as shown in the right part of Figure~\ref{model}, we introduce a lightweight language model $\theta^{\text{slm}}$ that maps $z_s$ to acoustic instructions $I_{emo}$. 
For example, ``\textit{The voice emotion should be full of playfulness... with an exaggerated tone...}".
This mapping ensures context-rich emotional intent is compressed into executable control signals. 

\noindent\textbf{Training with Instruction Control.} 
The talker is trained using a sequence modeling approach that concatenates instruction tokens, text tokens, and speech tokens. 
Training instructions are derived from comprehensive metadata extracted from large-scale speech corpora via open-source analysis models. 
This metadata-driven approach enables the talker to learn disentangled emotional style representations, treating $I_{emo}$ as a conditional prior that guides autoregressive acoustic token generation.

\noindent\textbf{Strategy-Driven Synthesis.} 
During inference, the talker operates under the instruction of the reasoned strategy. 
By conditioning synthesis on $I_{emo}$, we enforce rigorous intent and emotion consistency, ensuring the generated speech is not merely a textual reading but a holistic performance where prosodic features dynamically reflect the thinker's deep reasoning. 
Consequently, we complete the causal chain in Perception–Reasoning–Expression:
\begin{equation}
    % \begin{aligned}  % [l] 表示公式块整体左对齐，等号仍对齐
    I_{emo} = \theta^{\text{slm}}(z_s), Z = \theta^{\text{Thinker}}(\mathcal{M}), \hat{Y} = \theta^{\text{Talker}}(I_{\text{emo}}, z_t).
    % \end{aligned}
\end{equation}

\section{Experiments}

\textbf{Datasets.}
We utilize open-source MEU datasets, including MEIJU~\cite{MEIJU} and MER25~\cite{mer25}, as well as open-source MED datasets MultiDialog~\cite{multidialog} and the MELD~\cite{meld} training set. 
All open-source data are uniformly processed using our proposed data processing pipeline.
In addition, we incorporate privately processed movie data, which contributes to approximately 500 hours. The private data contains both Chinese and English dialogue.
In stage 1, the model is trained on the open-source dataset. 
In stage 2, we use all the datasets mentioned previously. 
Notably, for open-source datasets without responses, we utilize the synthetic dialogue responses generated by our data pipeline. 
When finetuning the talker, we use all annotated speech data and an additional 3000 hours of private high-quality TTS data.
The test set strictly follows our proposed EmoOmniEval benchmark.

\textbf{Implementation Details.}
We adopt Qwen2.5-Omni-7B as the base model for all experiments. 
The visual encoder and audio encoder of EmoOmni are frozen throughout the training phase.
All training experiments are executed on 16 NVIDIA A800 GPUs with mixed-precision enabled.
In the stage 1, the thinker is fully finetuned for 1 epoch with a total batch size of 128 and an initial learning rate of $1 \times 10^{-4}$.
In stage 2, we further finetune the thinker module using LoRA for two additional epochs, with a total batch size of 96 and a learning rate of $1 \times 10^{-4}$. The LoRA configuration employs a rank of 8, an alpha value of 32, and a dropout rate of 0.05.
When we finetune the talker, we use a total batch size of 128 for three epochs while maintaining a lower learning rate of $1 \times 10^{5}$.
In all experiments, we employ a cosine learning rate scheduler with 3,000 warm-up steps. 
All inputs are preprocessed using the Qwen2.5-Omni's processor to ensure compatibility with the base model.

\begin{table*}[htbp]

  \centering
  \caption{Comparison of model performance on EmoOmniEval, which includes two sets: MELD and ch-sims-v2. $^*$ denotes that the original model cannot synthesize speech and we therefore use EmoOmni-Talker to generate speech from its textual output for fair comparison.
  ``Param" indicates the parameters of the models.
  The best and second-best results are shown in bold and underlined, respectively.
  }
  % * 表示模型无法生成音频，我们使用豆包TTS做instructTTS.
% ^ 表示模型没有thinking过程，我们使用Gemini根据视频生成的strategy，来计算
  % \footnotesize
  \label{table1}
  \resizebox{\linewidth}{!}{
  % 调整字体大小以适应页面宽度，如果不需要可去掉 \small
  \setlength{\tabcolsep}{4pt} % 稍微减小列间距
\begin{tabular}{@{}l@{\hspace{3pt}}cccccccccccccc@{}}
% \begin{tabular}{lccllllllllllll}
  \toprule
  \multirow{2}{*}{Model/Metrics} & \multirow{3}{*}{\shortstack{Inference\\Mode}} & \multirow{2}{*}{Param} & \multicolumn{6}{c}{MELD} & \multicolumn{6}{c}{ch-sims-v2} \\
  % \cmidrule(lr){4-9} \cmidrule(lr){10-15}
  \cmidrule(lr){4-5} \cmidrule(lr){6-8} \cmidrule(lr){9-9} \cmidrule(lr){10-11} \cmidrule(lr){12-14} \cmidrule(lr){15-15}
  & & & VS-RES$\uparrow$ & VS-RC$\uparrow$ & VT-EA$\uparrow$ & VT-RES$\uparrow$ & VT-RC$\uparrow$ & IF$\uparrow$ & VS-RES$\uparrow$ & VS-RC$\uparrow$ & VT-EA$\uparrow$ & VT-RES$\uparrow$ & VT-RC$\uparrow$ & IF$\uparrow$ \\
  \midrule
  Gemini2.5Pro~\cite{gemini2.5}                    & -     & - & 1.42$^*$ & 1.60$^*$ & 1.97 & 1.98 & 1.86 & 1.29$^*$ & 1.87$^*$ & 1.92$^*$ & 2.00 & 2.00 & 1.99 & 1.60* \\
  \midrule
  Intern-S1~\cite{Interns1} & thinking           & 9B    & 0.52$^*$ & 0.62$^*$ & 0.45 & 0.42 & 0.49 & 1.16$^*$ & 1.05$^*$ & 1.27$^*$ & 0.76 & 1.03 & 1.18 & 1.33$^*$ \\
  MiniCPM-o2.6~\cite{minicpm} & no-thinking    & 8B    & 0.56 & 0.82 & - & - & 0.90 & - & 0.60 & 0.89 & - & - & 1.01 & - \\
  Qwen3Omni-Instruct~\cite{qwen3omni} & no-thinking & 30B & 0.91 & 1.25 & - & - & 1.22 & - & 1.43 & 1.60 & - & - & 1.59 & - \\
  Qwen3Omni-Thinking~\cite{qwen3omni} & thinking  & 30B   & \underline{1.28}$^*$ & \underline{1.34}$^*$ & \underline{1.32} & \underline{1.57} & \underline{1.47} & \textbf{1.34}$^*$ & \textbf{1.81}$^*$ & \textbf{1.91}$^*$ & \underline{1.56} & \underline{1.86} & \textbf{1.89} & \textbf{1.70}$^*$ \\
  Qwen2.5Omni~\cite{qwen2.5omni} & no-thinking      & 7B    & 0.93 & 1.18 & - & - & 1.07 & - & 1.20 & 1.46 & - & - & 1.34 & - \\
  EmoOmni & thinking    & 7B    & \textbf{1.36} & \textbf{1.40} & \textbf{1.93} & \textbf{1.95} & \textbf{1.53} & \underline{1.26} & \underline{1.67} & \underline{1.75} & \textbf{1.97} & \textbf{1.99} & \underline{1.81} & \underline{1.57} \\
  \bottomrule
\end{tabular}
}
\end{table*}

\textbf{Comparison Models.}
We compare our approach against several SOTA Omni-LLMs that are currently deployable by us: 
Qwen2.5-Omni-7B~\cite{qwen2.5omni}, Qwen3-Omni-30B-A3B-Instruct~\cite{qwen3omni}, Qwen3-Omni-30B-A3B-Thinking~\cite{qwen3omni}, Intern-S1-9B~\cite{Interns1}, and MiniCPM-o2.6-8B~\cite{minicpm}.
Notably, Intern-S1 and Qwen3-Omni-Thinking do not natively support speech generation and thus cannot be directly assessed on the VS dimension. 
To enable a fair comparison, we convert their thinking into TTS instructions and use EmoOmni-Talker to synthesize speech from their generated text responses.
Conversely, Qwen2.5-Omni, Qwen3-Omni-Instruct, and MiniCPM-o2.6 lack explicit thinking output, so for these models we do not report the score of VT-EA, VT-RES and IF.

\textbf{Evaluation Metrics.}
We include all the evaluation metrics defined in EmoOmniEval.
To assess speech intelligibility, following SeedTTS~\cite{seedtts}, we utilize
Whisper~\cite{whisper} to measure the Word-Error-Rate (WER) for
the English test set , and
Paraformer~\cite{paraformerv2} to measure Character-Error-Rate (CER) for
the Chinese test set of the generated speech. 
We use UTMOS~\cite{utmosv2} to assess the generated speech quality.
Additionally, we conduct subjective experiments to assess the effectiveness of the LLM-as-a-Judge. Comprehensive descriptions of the experimental setup and corresponding results are provided in the Appendix~\ref{app_sec_speech_wer_subjective} and Appendix~\ref{app_sec_subjective_eval_VS}.

\subsection{Experiments Results}
% Overall Superiority and the Impact of Ecological Grounding. 

\textbf{Overall.}
The quantitative results presented in Table~\ref{table1} demonstrate the results on our EmoOmniEval. Overall, considering VS-RES and VS-RC, EmoOmni consistently outperforms all models with comparable parameter scale and especially Qwen3Omni-Instruct. EmoOmni also achieves comparable performance with Qwen3Omni-Thinking.
When directly compared with its backbone model Qwen2.5-Omni-7B, EmoOmni exhibits consistent and substantial improvements across all metrics.
These results underscore the effectiveness of our explicitly decoupled Perception–Reasoning–Expression architecture and the reasoning-aware speech generation framework.

\textbf{Textual Analysis.}
Moreover, models equipped with thinking mechanisms consistently outperform their non-thinking counterparts on VT-RC and VT-RES metrics, highlighting that internal thinking processes serves as a crucial intermediate step for accurate emotional intent inference before expression.
On the VT-EA and VT-RC metrics, EmoOmni delivers great results on both metrics, reflecting its robust capability in jointly mastering multimodal emotional understanding and coherent, contextually appropriate dialogue generation.
Nevertheless, despite achieving relatively strong performance in the VT-EA (1.93) and VT-RES (1.95), EmoOmni’s final textual response score in VT-RC, only reaches 1.53. 
We attribute this performance gap primarily to two interrelated limitations: (1) the inherent capacity constraints of 7B parameters, and (2) the limited capabilities of the underlying base model.
We also observe that EmoOmni and the Qwen series perform better on the Chinese ch-sims-v2 testsets.
We attribute this tendency to differences in pretraining data distribution and linguistic coverage.

\textbf{IF Analysis.}
From the results, we can observe that the relative trends of the IF score are consistent across the two datasets.
However, it is evident that the IF metric exhibits considerable variation.
We assume that such variation arises from the existence of multiple variables, including the distinct semantic content and instruction designs of each model.
Furthermore, when using Gemini for evaluation and scoring, the semantic information conveyed by speech influences the final scores, thereby contributing to the observed variation.
Nevertheless, a higher IF score serves as indirect evidence that the generated speech exhibits better alignment with the corresponding text and instruction pairs.
This conclusion can be further validated by the results presented in Table~\ref{abl:different_talker}. Under strictly controlled variable conditions, the IF scores are reliable.

\textbf{Comparison of Thinker.}
Models marked with $^*$ differ only in their thinking modules, ensuring a fair comparison of textual reasoning quality.
Under this controlled setting, EmoOmni significantly outperforms Intern-S1-9B and achieves performance comparable to Qwen3Omni-Thinking, demonstrating the exceptional effectiveness of our thinker module.
Compared with end-to-end models including MiniCPM-o2.6 and Qwen3Omni-Instruct, EmoOmni demonstrates significantly stronger acoustic expression quality in multimodal emotional dialogue tasks.
This highlights the limitation of directly mapping multimodal inputs to speech without an explicit reasoning stage, which often leads to misaligned speech generation.

Although some models achieve competitive performance on VT metrics, their VS results lag behind significantly, indicating a semantic–acoustic emotion gap.
EmoOmni effectively narrows this gap, suggesting better alignment between internal emotional understanding and external expression.

\subsection{Ablation Studies}

\subsubsection{Variants of EmoOmni-Thinker}

\begin{table}[t]
  \centering
  \footnotesize
  \caption{Ablation study of each component within EmoOmni-Thinker conducted on ch-sims-v2 in EmoOmniEval.
   The best and second-best results are shown in bold and underlined, respectively.
  }
\label{emoomnithinker}
  \resizebox{\linewidth}{!}{
\begin{tabular}{@{}l@{\hspace{1pt}}c@{\hspace{3pt}}c@{\hspace{3pt}}c@{\hspace{3pt}}c@{\hspace{3pt}}c@{}}
  \toprule
  Model/Metrics & VS-RES$\uparrow$ & VS-RC$\uparrow$ & VT-EA$\uparrow$ & VT-RES$\uparrow$ & VT-RC$\uparrow$ \\
  \midrule
  EmoOmni                             & & & & & \\
  \quad w/ E-CoT                       & \textbf{1.67} & \textbf{1.75} & \underline{1.97} & \textbf{1.99} & \textbf{1.81} \\
  \quad \quad w/o Emotion Analysis  & 1.53 & 1.61 & -    & 1.58 & 1.62 \\
  \quad \quad w/o User Intent       & 1.63 & \underline{1.70} & 1.96 & 1.93 & 1.73 \\
  \quad \quad w/o Response Strategy & -    & -    & 1.97 & -    & 1.57 \\
  \quad \quad w ASR                 & 1.59 & 1.65 & 1.75 & 1.76 & 1.69 \\
  \quad w/o E-CoT                     & -    & -    & -    & -    & 1.36 \\
  \quad w/o Multitask                 & 1.65 & \textbf{1.75} & 1.88 & 1.89 & 1.67 \\
  \quad \quad w/o Stage1              & 1.64 & 1.66 & 1.67 & 1.68 & 1.68 \\
  \quad w/ SFT                         & \underline{1.66} & \textbf{1.75} & \textbf{1.98} & \underline{1.98} & \underline{1.77} \\
  \quad w/o Real-World Data           & 1.61 & 1.64 & 1.85 & 1.88 & 1.66 \\
  \bottomrule
\end{tabular}
  }
\end{table}

To comprehensively assess the contribution of each component within EmoOmni, we conducted a series of ablation experiments. These experiments as shown in Table~\ref{emoomnithinker} focus on disentangling the effects of the E-CoT, training strategies and training data.

\textbf{Impact of E-CoT:}
We first investigate the necessity of each component in the reasoning chain. Details of the E-CoT setups are provided in Appendix~\ref{app_sec_ecot_structure}.
Removing the multimodal emotion analysis module leads to incorrect emotional judgments in complex scenarios, degrading both response strategy and content.
Removing the user intent module slightly affects the model’s ability to understand user intentions.
Crucially, removing the Response Strategy node not only reduces VT-RC score, but also eliminates the generation of acoustic instructions, breaking the connection to the talker.
Interestingly, explicitly adding ASR transcripts as input harms performance, likely because the model shifts from multimodal perception to unimodal semantic analysis.
To isolate the effect of our proposed E-CoT framework, we trained a variant using the same dialogue data but without E-CoT. The results show a significant drop: VT-RC score falls from 1.84 to 1.36, which is only slightly better than the vanilla Qwen2.5-Omni.

\textbf{Impact of Training Strategy:} 
Removing the multi-task during stage 2 results in a slight degradation in VT-EA, which subsequently propagates errors to downstream generation tasks. 
Based on the results, removing stage 1 training leads to significant drops, confirming that the model must learn ``how to see" before learning ``how to reason". 
The ablation of our training strategy highlights the importance of the Perceptual Grounding phase.

\textbf{Efficacy of Data:} 
Regarding data ecology, removing the real-world dialogue data during training results in a substantial performance decline across all metrics. This validates our hypothesis that real-world social data in complex scenarios is indispensable for modeling the nuanced dynamics of human interaction, which synthetic or academic datasets fail to capture.

\subsubsection{Different Talkers}

\begin{table}[t]
\centering
\caption{
Ablation study on EmoOmni's performance when paired with different talkers conducted on ch-sims-v2 in EmoOmniEval.
$^*$ denotes that the original model lacks instruction control capability, we directly evaluate IF score between the generated speech and the fixed instruction.
The best results are shown in bold.
}
\label{abl:different_talker}
\resizebox{\linewidth}{!}{
% \footnotesize
\begin{tabular}{@{}l@{\hspace{1pt}}c@{\hspace{3pt}}c@{\hspace{3pt}}c@{\hspace{3pt}}c @{\hspace{3pt}}c@{}
}
\toprule
Model/Metrics & VS-RES$\uparrow$ & VS-RC$\uparrow$ & IF$\uparrow$ & WER$\downarrow$(\%) & UTMOS$\uparrow$ \\
\midrule
EmoOmni-Thinker \\
\quad w/ EmoOmni-Talker   & \textbf{1.67} & \textbf{1.75} & \textbf{1.57} & 4.72 & \textbf{2.69} \\
\quad w/ VoiceSculptor~\cite{VoiceSculptor}    & 1.64 & 1.71 & 1.48     & 6.62 & 2.67 \\
\quad w/ Llasa-3B~\cite{llasa}          & 1.61 & 1.71 & 1.30$^*$ & \textbf{4.45} & 2.73 \\
% \midrule
% \quad w Seed-tts-2.0     & 1.70 & 1.77 & 1.60 & 1.73 & 3.22 \\
% \quad w Qwen3-TTS             & \textbf{1.76} & \textbf{1.79} & \textbf{1.81} & – & – \\
\bottomrule
\end{tabular}
}
\end{table}

%先分析固定thinker 使用不同talker的区别。
From Table~\ref{abl:different_talker}, we observe that when using the same EmoOmni-Thinker, text and instructions, different talkers lead to noticeably different performance in VS-RES and VS-RC. This variation closely correlates with the TTS system's IF score. 
Our finetuned EmoOmni-Talker demonstrates better control over emotional instructions than the original VoiceSculptor, resulting in an improved IF score and, therefore, improved performance in VS-RES from 1.64 to 1.67.
Moreover, the comparison with LLaSA-3B further shows the effectiveness of incorporating instruction-based control in the speech synthesis for expressive output.

% Encouragingly, our EmoOmni-Talker approaches seed-tts-2.0's~\footnote{\url{https://www.volcengine.com/docs/6561/1329505?lang=zh}} performance closely, despite being a research-oriented model.
% Notably, results of seed-tts-2.0 highlight that the talker currently imposes a significant bottleneck on the expressive capacity of multimodal dialogue LLMs.
% Notably, the recently released Qwen3-TTS achieves VS-RES score of 1.76, highlighting that the Talker currently imposes a significant bottleneck on the expressive capacity of multimodal dialogue LLMs.
% Unfortunately, Qwen3-TTS is released too late for us to integrate it as our Talker in this work.

\section{Conclusion and Future Work}
\label{conclusion}
This paper proposes EmoOmni, which achieves emotionally intelligent interaction in multimodal dialogue.
Specifically, through the E-CoT and EmoOmni-Talker,  EmoOmni effectively bridges the perception and expression, ensuring the final response is semantically and emotionally aligned with the context.
To address data scarcity and assess the emotional multimodal dialogue task, we construct the EmoOmniPipe and EmoOmniEval.
Experiments show that EmoOmni-7B achieves comparable performance with Qwen3Omni-30B-A3B-Thinking under the same talker.

Future work will explore scaling EmoOmni on stronger omni-modal and TTS foundation models to achieve higher performance.
In addition, we aim to establish a fully end-to-end trainable framework that unifies multimodal perception and expression.
Moreover, extending EmoOmni to real-time, full-duplex interaction remains a promising direction, enabling continuous perception and emotionally aligned response generation in multi-turn interactive scenarios.

\newpage
\section*{Impact Statement}

\textbf{Social Impact.} 
Emotion plays an important role in human communication, conveying human intentions and deep thoughts. 
This paper introduces EmoOmni to enhance emotional intelligence in Omni-Modal Large Language Models. By bridging perception and expression through the Emotional Chain-of-Thought mechanism, our framework improves the naturalness of human-computer interaction. This technology is particularly beneficial for non-sensitive applications such as virtual companionship, interactive entertainment, and personalized education systems where emotional nuance is essential.

\textbf{Ethics Statement.}
The EmoOmniPipe constructs training data from publicly available movies, TV series, and open-source datasets. We do not collect private user information during this process. 
Therefore, we do not collect new data and we just re-annotate and filter existing data use the existing tools. 
To ensure data safety, we apply rigorous automated filters to remove toxic and incoherent content as detailed in the Appendix~\ref{app_sec_filter}. Furthermore, the annotation relies on advanced foundation models instead of human annotators, which avoids ethical concerns related to labor exploitation.
Therefore, no ethical concerns are raised in this paper.

\textbf{Proper Use and Risks.} Since our model learns from dramatic media, it may exhibit biases or generate factually incorrect content known as hallucinations. Additionally, the capability to generate emotionally expressive speech carries risks of misuse for emotional manipulation. We restrict the use of EmoOmni to academic research and strictly prohibit commercial applications. Users must not utilize this model in high-stakes scenarios involving medical advice, legal decisions, or deceptive activities.

\nocite{langley00}

\bibliography{example_paper}

@inproceedings{meld,
  author       = {Soujanya Poria and
                  Devamanyu Hazarika and
                  Navonil Majumder and
                  Gautam Naik and
                  Erik Cambria and
                  Rada Mihalcea},
  title        = {{MELD:} {A} Multimodal Multi-Party Dataset for Emotion Recognition
                  in Conversations},
  booktitle    = {Proc.{ACL}},
  pages        = {527--536},
  year         = {2019}
}

@article{qwen3omni,
  author       = {Jin Xu and
                  Zhifang Guo and
                  Hangrui Hu and
                  Yunfei Chu and
                  Xiong Wang and
                  Jinzheng He and others},
  title        = {Qwen3-Omni Technical Report},
  journal      = {CoRR},
  volume       = {abs/2509.17765},
  year         = {2025},
}

@article{qwen2.5omni,
  author       = {Jin Xu and
                  Zhifang Guo and
                  Jinzheng He and
                  Hangrui Hu and
                  Ting He and
                  Shuai Bai and
                  Keqin Chen and
                  Jialin Wang and
                  Yang Fan and
                  Kai Dang and
                  Bin Zhang and
                  Xiong Wang and
                  Yunfei Chu and
                  Junyang Lin},
  title        = {Qwen2.5-Omni Technical Report},
  journal      = {CoRR},
  volume       = {abs/2503.20215},
  year         = {2025},
}

@inproceedings{meiju,
  author       = {Rui Liu and
                  Xiaofen Xing and
                  Zheng Lian and
                  Haizhou Li and
                  Bj{\"{o}}rn W. Schuller and
                  Haolin Zuo},
  title        = {{MEIJU} - The 1st Multimodal Emotion and Intent Joint Understanding
                  Challenge},
  booktitle    = {Proc.{ICASSP}},
  pages        = {1--2},
  publisher    = {{IEEE}},
  year         = {2025},
}

@inproceedings{multidialog,
  author       = {Se Jin Park and
                  Chae Won Kim and
                  Hyeongseop Rha and
                  Minsu Kim and
                  Joanna Hong and
                  Jeong Hun Yeo and
                  Yong Man Ro},
  editor       = {Lun{-}Wei Ku and
                  Andre Martins and
                  Vivek Srikumar},
  title        = {Let's Go Real Talk: Spoken Dialogue Model for Face-to-Face Conversation},
  booktitle    = {Proc.{ACL}},
  pages        = {16334--16348},
  publisher    = {Association for Computational Linguistics},
  year         = {2024},
}

@article{mer25,
  author       = {Zheng Lian and
                  Rui Liu and
                  Kele Xu and
                  Bin Liu and
                  Xuefei Liu and
                  Yazhou Zhang and
                  Xin Liu and
                  Yong Li and
                  Zebang Cheng and
                  Haolin Zuo and
                  Ziyang Ma and
                  Xiaojiang Peng and
                  Xie Chen and
                  Ya Li and
                  Erik Cambria and
                  Guoying Zhao and
                  Bj{\"{o}}rn W. Schuller and
                  Jianhua Tao},
  title        = {{MER} 2025: When Affective Computing Meets Large Language Models},
  journal      = {CoRR},
  volume       = {abs/2504.19423},
  year         = {2025},
}

@article{Interns1,
  author       = {Lei Bai and
                  Zhongrui Cai and
                  Yuhang Cao and
                  Maosong Cao and
                  Weihan Cao and
                  Chiyu Chen and others},
  title        = {Intern-S1: {A} Scientific Multimodal Foundation Model},
  journal      = {CoRR},
  volume       = {abs/2508.15763},
  year         = {2025},
}

@misc{minicpm,
  title={Minicpm-o 2.6: A gpt-4o level mllm for vision, speech, and multimodal live streaming on your phone},
  author={Team, OpenBMB MiniCPM-o},
  year={2025}
}

@article{VoiceSculptor,
  title={VoiceSculptor: Your Voice, Designed By You},
  author={Hu, Jingbin and Chen, Huakang and Ma, Linhan and Guo, Dake and Zhan, Qirui and Li, Wenhao and Zhang, Haoyu and Xia, Kangxiang and Zhang, Ziyu and Tian, Wenjie and others},
  journal      = {CoRR},   volume       = {abs/2601.10629},
  year={2026}
}

@article{llasa,
  title={Llasa: Scaling train-time and inference-time compute for llama-based speech synthesis},
  author={Ye, Zhen and Zhu, Xinfa and Chan, Chi-Min and Wang, Xinsheng and Tan, Xu and Lei, Jiahe and Peng, Yi and Liu, Haohe and Jin, Yizhu and Dai, Zheqi and others},
  journal      = {CoRR},   volume       = {abs/2502.04128},
  year={2025}
}

@article{gemini2.5,
  title={Gemini 2.5: Pushing the frontier with advanced reasoning, multimodality, long context, and next generation agentic capabilities},
  author={Comanici, Gheorghe and Bieber, Eric and Schaekermann, Mike and Pasupat, Ice and Sachdeva, Noveen and others},
  journal      = {CoRR},   volume       = {abs/2507.06261},
  year={2025}
}

@inproceedings{unimse,
  author       = {Guimin Hu and
                  Ting{-}En Lin and
                  Yi Zhao and
                  Guangming Lu and
                  Yuchuan Wu and
                  Yongbin Li},
  title        = {UniMSE: Towards Unified Multimodal Sentiment Analysis and Emotion
                  Recognition},
  booktitle    = {Proc.{EMNLP}},
  pages        = {7837--7851},
  year         = {2022},
}

@inproceedings{mmgcn,
  author       = {Jingwen Hu and
                  Yuchen Liu and
                  Jinming Zhao and
                  Qin Jin},
  title        = {{MMGCN:} Multimodal Fusion via Deep Graph Convolution Network for
                  Emotion Recognition in Conversation},
  booktitle    = {Proc.{ACL/IJCNLP}},
  pages        = {5666--5675},
  year         = {2021},
}

@inproceedings{emotion-llama,
  author       = {Zebang Cheng and
                  Zhi{-}Qi Cheng and
                  Jun{-}Yan He and
                  Kai Wang and
                  Yuxiang Lin and
                  others},
  title        = {Emotion-LLaMA: Multimodal Emotion Recognition and Reasoning with Instruction
                  Tuning},
  booktitle    = {Proc.{NeurIPS}},
  year         = {2024},
}

@inproceedings{mosear,
  title={Benchmarking and Bridging Emotion Conflicts for Multimodal Emotion Reasoning},
  author={Han, Zhiyuan and Zhu, Beier and Xu, Yanlong and Song, Peipei and Yang, Xun},
  booktitle={Proc.{ACM MM}},
  year={2025}
}

@article{humanomni,
  author       = {Jiaxing Zhao and
                  Qize Yang and
                  Yixing Peng and
                  Detao Bai and
                  Shimin Yao and
                  Boyuan Sun and
                  Xiang Chen and
                  Shenghao Fu and
                  Weixuan chen and
                  Xihan Wei and
                  Liefeng Bo},
  title        = {HumanOmni: {A} Large Vision-Speech Language Model for Human-Centric
                  Video Understanding},
  journal      = {CoRR},
  volume       = {abs/2501.15111},
  year         = {2025},
}

@article{indextts2,
  author       = {Siyi Zhou and
                  Yiquan Zhou and
                  Yi He and
                  Xun Zhou and
                  Jinchao Wang and
                  Wei Deng and
                  Jingchen Shu},
  title        = {IndexTTS2: {A} Breakthrough in Emotionally Expressive and Duration-Controlled
                  Auto-Regressive Zero-Shot Text-to-Speech},
  journal      = {CoRR},
  volume       = {abs/2506.21619},
  year         = {2025},
}

@inproceedings{AffectGPT,
  author       = {Zheng Lian and
                  Haoyu Chen and
                  Lan Chen and
                  Haiyang Sun and
                  Licai Sun and
                  Yong Ren and
                  Zebang Cheng and
                  Bin Liu and
                  Rui Liu and
                  Xiaojiang Peng and
                  Jiangyan Yi and
                  Jianhua Tao},
  title        = {AffectGPT: {A} New Dataset, Model, and Benchmark for Emotion Understanding
                  with Multimodal Large Language Models},
  booktitle    = {Proc.{ICML}},
  publisher    = {OpenReview.net},
  year         = {2025},
}

@article{merbench,
  author       = {Zheng Lian and
                  Licai Sun and
                  Yong Ren and
                  Hao Gu and
                  Haiyang Sun and
                  Lan Chen and
                  Bin Liu and
                  Jianhua Tao},
  title        = {MERBench: {A} Unified Evaluation Benchmark for Multimodal Emotion
                  Recognition},
  journal      = {CoRR},
  volume       = {abs/2401.03429},
  year         = {2024},
}

@article{stepaudio2,
  author       = {Boyong Wu and
                  Chao Yan and
                  Chen Hu and
                  Cheng Yi and
                  Chengli Feng and
                  Fei Tian and others},
  title        = {Step-Audio 2 Technical Report},
  journal      = {CoRR},
  volume       = {abs/2507.16632},
  year         = {2025},
}

@article{instructTTS,
  author       = {Dongchao Yang and
                  Songxiang Liu and
                  Rongjie Huang and
                  Chao Weng and
                  Helen Meng},
  title        = {InstructTTS: Modelling Expressive {TTS} in Discrete Latent Space With
                  Natural Language Style Prompt},
  journal      = {{IEEE} {ACM} Trans. Audio Speech Lang. Process.},
  volume       = {32},
  pages        = {2913--2925},
  year         = {2024},
}

@article{mingomni,
  author       = {Inclusion AI and
                  Biao Gong and
                  Cheng Zou and
                  Chuanyang Zheng and
                  Chunluan Zhou and
                  Canxiang Yan and others},
  title        = {Ming-Omni: {A} Unified Multimodal Model for Perception and Generation},
  journal      = {CoRR},
  volume       = {abs/2506.09344},
  year         = {2025},
}

@article{seedtts,
  author       = {Philip Anastassiou and
                  Jiawei Chen and
                  Jitong Chen and
                  Yuanzhe Chen and
                  Zhuo Chen and
                  Ziyi Chen and others},
  title        = {Seed-TTS: {A} Family of High-Quality Versatile Speech Generation Models},
  journal      = {CoRR},
  volume       = {abs/2406.02430},
  year         = {2024},
}

@inproceedings{whisper,
  author       = {Alec Radford and
                  Jong Wook Kim and
                  Tao Xu and
                  Greg Brockman and
                  Christine McLeavey and
                  Ilya Sutskever},
  title        = {Robust Speech Recognition via Large-Scale Weak Supervision},
  booktitle    = {Proc.{ICML}},
  volume       = {202},
  pages        = {28492--28518},
  year         = {2023},
}

@article{paraformerv2,
  author       = {Keyu An and
                  Zerui Li and
                  Zhifu Gao and
                  Shiliang Zhang},
  title        = {Paraformer-v2: An improved non-autoregressive transformer for noise-robust
                  speech recognition},
  journal      = {CoRR},
  volume       = {abs/2409.17746},
  year         = {2024},
}

@article{qwen3,
  author       = {An Yang and
                  Anfeng Li and
                  Baosong Yang and
                  Beichen Zhang and
                  Binyuan Hui and
                  Bo Zheng and others},
  title        = {Qwen3 Technical Report},
  journal      = {CoRR},
  volume       = {abs/2505.09388},
  year         = {2025},
}

@article{saber,
  author={Zhixian Zhao and Wenjie Tian and Xiaohai Tian and Jun Zhang and Lei Xie},
  title={Integrating Fine-Grained Audio-Visual Evidence for Robust Multimodal Emotion Reasoning}, 
  journal      = {CoRR},
  volume       = {abs/2601.18321},
  year         = {2026},
}

@article{melbandroformer,
  author       = {Ju{-}Chiang Wang and
                  Wei Tsung Lu and
                  Minz Won},
  title        = {Mel-Band RoFormer for Music Source Separation},
  journal      = {CoRR},
  volume       = {abs/2310.01809},
  year         = {2023},
}

@article{gpt4o,
  author       = {Aaron Hurst and
                  Adam Lerer and
                  Adam P. Goucher and
                  Adam Perelman and
                  Aditya Ramesh and
                  Aidan Clark and others},
  title        = {GPT-4o System Card},
  journal      = {CoRR},
  volume       = {abs/2410.21276},
  year         = {2024},
}

@article{labeldata_dialogue1,
  author       = {Ronghao Lin and
                  Shuai Shen and
                  Weipeng Hu and
                  Qiaolin He and
                  Aolin Xiong and
                  Li Huang and
                  Haifeng Hu and
                  Yap{-}Peng Tan},
  title        = {{E3RG:} Building Explicit Emotion-driven Empathetic Response Generation
                  System with Multimodal Large Language Model},
  journal      = {CoRR},
  volume       = {abs/2508.12854},
  year         = {2025},
}

@article{mllm_1,
  author       = {Ronghao Lin and
                  Shuai Shen and
                  Weipeng Hu and
                  Qiaolin He and
                  Aolin Xiong and
                  Li Huang and
                  Haifeng Hu and
                  Yap{-}Peng Tan},
  title        = {{E3RG:} Building Explicit Emotion-driven Empathetic Response Generation
                  System with Multimodal Large Language Model},
  journal      = {CoRR},
  volume       = {abs/2508.12854},
  year         = {2025},
}

@article{mllm_2,
  author       = {Weiting Tan and
                  Jiachen Lian and
                  Hirofumi Inaguma and
                  Paden Tomasello and
                  Philipp Koehn and
                  Xutai Ma},
  title        = {Seeing is Believing: Emotion-Aware Audio-Visual Language Modeling
                  for Expressive Speech Generation},
  journal      = {CoRR},
  volume       = {abs/2508.16188},
  year         = {2025},
}

@inproceedings{utmosv2,
  title     = {The T05 System for The {V}oice{MOS} {C}hallenge 2024: Transfer Learning from Deep Image Classifier to Naturalness {MOS} Prediction of High-Quality Synthetic Speech},
  author    = {Baba, Kaito and Nakata, Wataru and Saito, Yuki and Saruwatari, Hiroshi},
  booktitle = {IEEE Spoken Language Technology Workshop (SLT)},
  year      = {2024},
}

@article{MMLA,
  author={Zhang, Hanlei and Li, Zhuohang and Zhu, Yeshuang and Xu, Hua and Wang, Peiwu and Zhu, Haige and Zhou, Jie and Zhang, Jinchao},
  title={Can Large Language Models Help Multimodal Language Analysis? MMLA: A Comprehensive Benchmark},
  journal      = {CoRR},
  volume       = {abs/2504.16427},
  year         = {2025},
}

@article{emobench-m,
  author       = {He Hu and
                  Yucheng Zhou and
                  Lianzhong You and
                  Hongbo Xu and
                  Qianning Wang and others},
  title        = {EmoBench-M: Benchmarking Emotional Intelligence for Multimodal Large
                  Language Models},
  journal      = {CoRR},
  volume       = {abs/2502.04424},
  year         = {2025},
}

@article{merllm,
  author={Li, Zinuo and Zhang, Xian and Guo, Yongxin and Bennamoun, Mohammed and Boussaid, Farid and Dwivedi, Girish and Gong, Luqi and Ke, Qiuhong},
  title={Watch and Listen: Understanding Audio-Visual-Speech Moments with Multimodal LLM},
  journal      = {CoRR},
  volume       = {abs/2505.18110},
  year         = {2025},
}

@article{merllm2,

  author={Yang, Qize and Bai, Detao and Peng, Yi-Xing and Wei, Xihan},
  title={Omni-emotion: Extending video mllm with detailed face and audio modeling for multimodal emotion analysis},
  journal      = {CoRR},
  volume       = {abs/2501.09502},
  year         = {2025},
}
\bibliographystyle{icml2026}

%%%%%%%%%%%%%%%%%%%%%%%%%%%%%%%%%%%%%%%%%%%%%%%%%%%%%%%%%%%%%%%%%%%%%%%%%%%%%%%
%%%%%%%%%%%%%%%%%%%%%%%%%%%%%%%%%%%%%%%%%%%%%%%%%%%%%%%%%%%%%%%%%%%%%%%%%%%%%%%
% APPENDIX
%%%%%%%%%%%%%%%%%%%%%%%%%%%%%%%%%%%%%%%%%%%%%%%%%%%%%%%%%%%%%%%%%%%%%%%%%%%%%%%
%%%%%%%%%%%%%%%%%%%%%%%%%%%%%%%%%%%%%%%%%%%%%%%%%%%%%%%%%%%%%%%%%%%%%%%%%%%%%%%
\newpage
\appendix
\onecolumn

\section{E-CoT Structure}
\label{app_sec_ecot_structure}
Our complete E-CoT framework consists of four core components: multimodal emotion analysis, user intent recognition, response strategy planning, and response content generation.
To validate the necessity of each component, we conducted a series of ablation experiments as follows.

\textbf{w/o Emotion Analysis}: The model pipeline is reduced to user intent recognition, response strategy planning, and response content generation, which helps verify the impact of emotional information on the final response.

\textbf{w/o User Intent}: The model pipeline is reduced to multimodal emotion analysis, response strategy planning, and response content generation, which helps validate the critical role of intent understanding in conversations.

\textbf{w/o Response Strategy}: The model pipeline is reduced to multimodal emotion analysis, user intent recognition, and response content generation, which helps evaluate the contribution of strategy planning to the logical consistency and adaptability of responses.

\textbf{w ASR}: ASR transcription is added before the original pipeline, forming an end-to-end flow of ASR text, multimodal emotion analysis, user intent recognition, response strategy planning and response content generation, which validates the framework's performance in speech input scenarios.

\section{WER and Subjective Metric}
\label{app_sec_speech_wer_subjective}

To assess the quality of the generated responses, we conduct a series of subjective listening experiments with human evaluators. Specifically, we evaluate each baseline model along three dimensions: Naturalness of the synthesized speech, measured by the Natural Mean Opinion Score (N-MOS) on a 5-point scale (1 = unnatural, 5 = completely natural).
Relevance and Coherence of the response in the visual-audio context, scored as the Mean Opinion Score for Video-Speech Relevance and Coherence (VS-RC-MOS).
Emotional Appropriateness of the response given the multimodal input, scored as the Mean Opinion Score for Video-Speech Response Emotional Strategy (VS-RES-MOS).
Both VS-RC-MOS and VS-RES-MOS are evaluated on a 2-point scale, following the LLM-as-a-Judge protocol but performed by human annotators to ensure alignment with human perception.
20 participants are recruited for the evaluation. Each participant was presented with audio-visual stimuli paired with model-generated responses and asked to rate them according to the above criteria. All reported scores include 95\% confidence intervals to reflect statistical reliability.

As shown in Table~\ref{app_table_speech_wer_subjective}, although EmoOmni does not achieve the best WER, it obtains the highest N-MOS score. This observation is expected, as WER mainly reflects transcription accuracy, whereas N-MOS directly evaluates perceptual naturalness, which is strongly influenced by prosody and expressiveness. 
EmoOmni explicitly optimizes emotional alignment in speech generation, leading to more natural speech even when minor WER are present.
\begin{table}[h]
  \centering
  \caption{Speech Generation Performance on ch2-sims-v2 Dataset}
  \label{app_table_speech_wer_subjective}
  \begin{tabular}{lcc}
    \toprule
    \multicolumn{3}{c}{ch2-sims-v2} \\
    \midrule
    Model/Metrics & WER$\downarrow$(\%) & N-MOS$\uparrow$ \\
    \midrule
    MiniCPM-o2.6 & 12.12 & 3.39 $\pm$ 0.29 \\
    Qwen3Omni-Instruct & 3.35 & 4.08 $\pm$ 0.37 \\
    Qwen2.5Omni & 1.17 & 3.96 $\pm$ 0.44 \\
     EmoOmni & 4.72 & 4.17 $\pm$ 0.21 \\
    \bottomrule
  \end{tabular}
\end{table}

\section{Subjective Metrics of VS-RES and VS-RC}
\label{app_sec_subjective_eval_VS}
Regarding the subjective human assessment metrics shown in Table~\ref{app_table_subjective_eval_VS}, including RES-MOS and RC-MOS, our EmoOmni achieves outcomes that are merely slightly lower than those of Qwen3Omni-Thinking. Importantly, the consistency between this trend and the findings in Table 1 serves as strong evidence for the effectiveness of our Gemini-based testing protocol.
\begin{table}[htbp]
  \centering
  \caption{Subjective Human Evaluation Results on ch2-sims-v2 Dataset}
  \label{app_table_subjective_eval_VS}
  \begin{tabular}{lcc}
    \toprule
    \multicolumn{1}{c}{Subjective} & \multicolumn{2}{c}{ch2-sims-v2} \\
    \midrule
    Model/Metrics & VS-RES-MOS$\uparrow$ & VS-RC-MOS$\uparrow$ \\
    \midrule
    Gemini2.5Pro & 1.78 $\pm$ 0.14 & 1.98 $\pm$ 0.11 \\
    Intern-S1 & 0.88 $\pm$ 0.22 & 1.01 $\pm$ 0.14 \\
    MiniCPM-o2.6 & 0.33 $\pm$ 0.07 & 0.46 $\pm$ 0.08 \\
    Qwen3Omni-Instruct & 1.30 $\pm$ 0.16 & 1.60 $\pm$ 0.14 \\
    Qwen3Omni-Thinking & 1.68 $\pm$ 0.12 & 1.95 $\pm$ 0.13 \\
    Qwen2.5Omni & 1.03 $\pm$ 0.09 & 1.30 $\pm$ 0.07 \\
    EmoOmni & 1.56 $\pm$ 0.10 & 1.81 $\pm$ 0.09 \\
    \bottomrule
  \end{tabular}
\end{table}

\section{Different Talkers}
\label{app_sec_differencetalker}
Notably, results of seed-tts-2.0~\footnote{\url{https://www.volcengine.com/docs/6561/1329505?lang=zh}} highlight that the talker currently imposes a significant bottleneck on the expressive capacity.
Encouragingly, our EmoOmni-Talker approaches seed-tts-2.0's performance closely, despite being a research-oriented model.
The recently released Qwen3-TTS~\footnote{\url{https://github.com/QwenLM/Qwen3-TTS}} achieves VS-RES score of 1.76, highlighting that the talker currently imposes a significant bottleneck on the expressive capacity of multimodal dialogue LLMs.
This further confirms that the talker component has consistently limited the model’s expressive performance.
This is essential for improving the quality of multimodal interaction within the Perception-Reasoning-Expression causal chain.
But Qwen3-TTS is released too late for us to integrate it as our talker in this work.
\begin{table}[h]
\centering
\caption{
Ablation study on EmoOmni's performance when paired with different talkers conducted on ch-sims-v2 in EmoOmniEval.
}
\label{app_table_differencetalker}
% \resizebox{\linewidth}{!}{
\footnotesize
\begin{tabular}{@{}l@{\hspace{3pt}} c@{\hspace{3pt}} c@{\hspace{3pt}} c@{\hspace{3pt}} c @{\hspace{3pt}}c @{}
}
\toprule
Model/Metrics & VS-RES$\uparrow$ & VS-RC$\uparrow$ & IF$\uparrow$ & WER$\downarrow$(\%) & UTMOS$\uparrow$ \\
\midrule
EmoOmni-Thinker \\
\quad w/ EmoOmni-Talker   & 1.67 & 1.75 & 1.57 & 4.72 & 2.69 \\
\quad w/ Seed-tts-2.0     & 1.70 & 1.77 & 1.60 & 1.73 & 3.22 \\
\quad w/ Qwen3-TTS        & 1.76 & 1.79 & 1.81 & 2.40 & 3.03 \\
\bottomrule
\end{tabular}
\end{table}

\section{Training Data Details}
\label{app_sec_trainingdata}

\begin{table}[htbp]
\centering
\caption{Overview of training data. ``Gemini" indicates the answer is generated by Gemini2.5Pro.}
\label{app_table_trainingdata}
\begin{tabular}{l c c c c c}
\toprule
\multirow{2}{*}{Dataset} & \multirow{2}{*}{Language}  & \multicolumn{2}{c}{Emotion Understanding Data} & \multicolumn{2}{c}{Emotion Dialogue Data} \\
\cmidrule(lr){3-4} \cmidrule(lr){5-6}
& & Scale & Source & Scale & Source \\
\midrule

MEIJU25 (CN)~\cite{MEIJU} & CN & 114.1k & TV         & 50.6k & Gemini \\
MEIJU25 (EN)~\cite{MEIJU} & EN & 85.9k & TV          & 39.1k & Gemini \\
MER25~\cite{mer25} & CN & 132.2k & TV                & 114.7k & Gemini \\
MELD~\cite{meld} & EN & 13.7k & TV                   & 3.1k & GroundTruth \\
MultiDialog~\cite{multidialog} & EN & 187.9k & Acted & 85.3k & GroundTruth \\
\midrule
Private Movies and TVs & CN/EN & 183.1k & Acted      & 91.6k & GroundTruth \\
\midrule
EmoOmni & CN/EN & 716.9k & Mixed & 384.4k & Mixed \\ % 示例值，可替换为真实数据
\bottomrule
\end{tabular}
\end{table}

% \section{EmoOmniEval Data Details}
% \label{app_sec_benchmarkdata}

% \begin{table}[htbp]
% \centering
% \caption{Overview of open-sourced multimodal emotion understanding datasets and emotion  dialogue datasets.}
% \label{app_table_benchmarkdata}
% % \setlength{\tabcolsep}{1.5pt} % 压缩列间距
% \begin{tabular}{l c c c c c}
% \toprule
% \multirow{2}{*}{Dataset} & \multirow{2}{*}{Language}  & \multicolumn{2}{c}{Emotion Understanding Data} & \multicolumn{2}{c}{Emotion Dialogue Data} \\
% \cmidrule(lr){3-4} \cmidrule(lr){5-6}
% & & Scale & Source & Scale & Source \\
% \midrule
% MEIJU25 (CN)~\cite{MEIJU} & CN & 114.1k & TV         & - & Gemini \\
% MEIJU25 (EN)~\cite{MEIJU} & EN & 85.9k & TV          & - & Gemini \\
% MER25~\cite{mer25} & CN & 132.2k & TV                & - & Gemini \\
% MELD~\cite{meld} & EN & 13.7k & TV                   & - & GT \\
% MultiDialog~\cite{multidialog} & EN & 187.9k & Acted & - & GT \\
% \midrule
% Private Movies and TVs & CN/EN & 187.9k & Acted      & - & GT \\
% \midrule
% \textbf{EmoOmni} & \textbf{CN/EN} & \textbf{600k} & \textbf{Mixed} & \textbf{800k} & \textbf{Mixed} \\ % 示例值，可替换为真实数据
% \bottomrule
% \end{tabular}
% \end{table}

\section{The Impact of Training Data}
\label{app_sec_number_of_trainingdata}

An ablation study was conducted on CH-SIMS-v2 to investigate the impact of varying the amount of task 1 training data during stage 2 multitask learning. 
As the amount of data changes shown in Table~\ref{app_table_number_of_trainingdata}, the model performance exhibits the following trend: optimal results are achieved when the amount of task 1 data is approximately 100 hours.
This phenomenon can be attributed to the limited scale of our task 2 data. When an excessive amount of Task 1 data is introduced, it may interfere with the model's ability to effectively learn the conversational task. Therefore, balancing the data ratio between multitasks is crucial for achieving optimal performance.

\begin{table}[htbp]
\centering
    \caption{The impact of data volume on multi-task in task 1 learning performance evaluated on the ch2-sims-v2 test set.}
\label{app_table_number_of_trainingdata}
\begin{tabular}{lccc}
\toprule
    Task1 Data/Hours & VT-EA$\uparrow$ & VT-RES$\uparrow$ & VT-RC$\uparrow$ \\\midrule
    0 & 1.88 & 1.89 & 1.67 \\
    100 & 1.97 & 1.99 & 1.81 \\
    500 & 1.97 & 1.96 & 1.71 \\
    1000 & 1.96 & 1.98 & 1.75 \\\bottomrule\end{tabular}

\end{table}

% \section{IF Score of Baselines}

% \begin{table}[htbp]
%   \centering
%   \caption{IF Performance of Different Models Based on Table~\ref{table1}.}
%   \label{tab:if_performance}
%   \begin{tabular}{lcc}
%     \toprule
%     Model/Test sets & MELD & ch2-sims-v2 \\
%     \midrule
%     Gemini2.5Pro & 1.29 & 1.60 \\
%     Intern-S1 & 1.16 & 1.33 \\
%     % MiniCPM-o 2.6 & 1.16* & -- \\
%     % qwen3omni-instruct & -- & -- \\
%     Qwen3Omni-Thinking & 1.34 & 1.70 \\
%     % qwen2.5omni & 1.34* & -- \\
%     \textbf{EmoOmni} & 1.26 & 1.57 \\
%     \bottomrule
%   \end{tabular}
% \end{table}

% From the results, we can observe that the relative trends of the IF score are consistent across the two datasets. However, when large models are scored from the IF perspective, the semantic content of the speech directly influences their scores. This also indirectly reflects the degree of alignment between the generated text and the given instruction.
% The results in Table~\ref{abl:different_talker} further validate that our IF metric is functioning as expected. The missing values in the table are due to the high number of variables involved in the experimental setup, which made it challenging to maintain consistent conditions across all test cases.

\section{Data Filter Prompt}
\label{app_sec_filter}

\subsection{Semantic Filter}
\label{appendix:semantic_filter}

\begin{tcolorbox}[colback=white, colframe=black, title=System Prompt, breakable]

\textbf{Role}

You are a top-tier data labeling expert specializing in selecting high-quality dialogue data for training Large Language Models. Your core task is to evaluate the internal logical coherence and information value of dialogue fragments.

\textbf{Task}

Analyze the dialogue fragment provided below and determine if it constitutes a logically coherent and meaningful unit of exchange. This unit will be used to train a large language model, so it needs to demonstrate natural language communication patterns.

\textbf{Core Evaluation Criteria}

\begin{enumerate}[nosep, leftmargin=*]
    \item \textbf{Logical Coherence (Most Important)}: Is every sentence in the dialogue a reasonable continuation, answer, or reaction to the previous context? Does the entire fragment revolve around one or more associated topics? Are there sudden, illogical topic jumps?
    \item \textbf{Dialogue Interaction}: Does the fragment contain at least one meaningful "Question \& Answer", "Request \& Response", or "Statement \& Comment" multi-turn interaction? A simple statement followed by another unrelated statement has low value.
    \item \textbf{Information Completeness}: Although the fragment may be part of a longer conversation, does it convey a relatively complete intent or information point? For example, a complete inquiry for directions, a brief argument, a handover of tasks, etc.
\end{enumerate}

\textbf{Data is Good (What we need)}

\begin{itemize}[nosep, leftmargin=*]
    \item Fragments containing multi-turn Q\&A.
    \item Characters discussing or arguing about a topic.
    \item Even if it starts with "So..." or "But...", as long as the subsequent dialogue is coherent, we consider it good.
    \item Even if the ending is an open-ended question (e.g., "How is the situation over there?"), as long as it logically follows the previous content, we consider it good.
\end{itemize}

\textbf{Data is Bad (To be filtered out)}

\begin{itemize}[nosep, leftmargin=*]
    \item Monologue: Only one person speaking continuously without interaction.
    \item Logical Break: No connection between sentences, appearing randomly spliced.
    \item Unclear Meaning: Dialogue content is too fragmented to understand its main idea. For example, containing only meaningless responses like "Um", "Ah", "Okay".
\end{itemize}

\textbf{Output Format}

Strictly provide your analysis results in the following JSON format without adding any additional explanation or text.

\begin{verbatim}
{
  "is_coherent_and_valuable": true or false,
  "confidence": "High/Medium/Low",
  "reason": "Briefly explain your judgment in one sentence. E.g., The fragment 
             revolves around an inquiry, contains multi-turn Q&A, and is 
             logically coherent. OR: The fragment only contains meaningless 
             responses and lacks informational value."
}
\end{verbatim}

% \textbf{Example 1 (Bad Data)}

% Dialogue Text:
% Speaker 1: Hahahaha.
% Speaker 2: Yeah.
% Speaker 1: Let's go.

% Output:
% \begin{verbatim}
% {
%   "is_coherent_and_valuable": false,
%   "confidence": "High",
%   "reason": "The fragment consists of meaningless interjections and a single 
%              instruction, lacking coherent conversational context and value."
% }
% \end{verbatim}

\end{tcolorbox}

\subsection{Toxic Filter}
\label{appendix:toxic_filter}

\begin{tcolorbox}[colback=white, colframe=black, title=Evaluation Prompt, breakable]

\textbf{Role}

You are a strict data quality cleaning expert. Your task is to audit whether the "Dialogue Strategy Analysis" contains "Interpretive Hallucinations".

\textbf{Task}

I will provide you with two parts:
\begin{enumerate}[nosep, leftmargin=*]
    \item \textbf{Dialogue Context}: The actual historical dialogue record.
    \item \textbf{Response Strategy Analysis}: The analysis generated by the model explaining why it responded in that way.
\end{enumerate}

You need to judge whether the reasons mentioned in the Response Strategy Analysis are faithful to the Dialogue Context.

\textbf{Definition of Hallucination}

In real dialogue data, sometimes the Real Response does not match the sentiment of the Context (e.g., the context is happy, but the response is negative). To forcibly explain this mismatch, the model might fabricate facts that do not exist in the context within the Response Strategy Analysis.

\textbf{If specific reasons such as "because I am tired," "because I am sick," or "because I had bad luck" appear in the strategy analysis, but are completely unmentioned in the context, this constitutes a severe hallucination.}

\textbf{Judgment Criteria}

\begin{enumerate}[nosep, leftmargin=*]
    \item \textbf{PASS}: The strategy analysis is entirely based on information in the context, or only makes reasonable inferences about the emotions of both parties (e.g., "Although the user is happy, the responder appears relatively cold"), without fabricating specific facts.
    \item \textbf{REJECT}: The strategy analysis introduces external information, fabricates character states (such as being tired, busy, sick) or events not present in the context to justify the response.
\end{enumerate}

If hallucination occurs (REJECT), please specify the reject\_condition. If passed, set reject\_condition to none.

\begin{enumerate}[nosep, leftmargin=*]
    \item \textbf{answer\_reason}: In the response, the 'responder' or 'I' fabricates personal states (such as fatigue, busyness, illness) or events that do not exist to defend the response.
    \item \textbf{context\_reason}: Fabrications involving elements other than 'me', such as fabricating the other party's state.
\end{enumerate}

\textbf{Additionally, please determine if this response is suitable for AI assistant training (i.e., is the dialogue reasonable, and does it avoid inappropriate negative emotional responses).} (suit)

\textbf{Output Format}

Please output only in JSON format, without other nonsense:

\begin{verbatim}
{
    "judgment": "PASS" or "REJECT",
    "reason": "Brief explanation of the reason",
    "reject_condition": "context_reason" or "answer_reason" or "none",
    "suit": "YES" or "NO"
}
\end{verbatim}

% \textbf{Now I will provide you with the data, please start the analysis:}

% Dialogue Context: \{Context\}. 
% Response Strategy Analysis: \{Model Analysis\}

\end{tcolorbox}

\section{Benchmark Prompt}
\label{benchmarkprompt}

\subsection{VT-RC Score Prompt}
\label{app:eval_prompt}

\begin{tcolorbox}[colback=white, colframe=black, title=Evaluation Prompt, breakable]

\textbf{Role}

You are an evaluation expert proficient in multimodal interaction and affective computing. Your task is to evaluate the performance of a conversational model with Emotional Chain-of-Thought (CoT) based on raw video input.

\textbf{Input Data}

\begin{enumerate}[nosep, leftmargin=*]
    \item \textbf{Video Input}: The raw video snippet input by the user. This serves as the context for the user conversation.
    \item \textbf{Model Response}: The response generated by the model, serving as the text reply to the current user.
\end{enumerate}

\textbf{Evaluation Task}

Based on the video content, please score (0-2 points) and critique the model's performance according to the following dimension.

\textit{Dimension 1: Response Content Relevance \& Logic}

\textbf{Goal}: Evaluate whether the text content of Model Response closely adheres to the semantic information in Video Input and whether the logic is coherent.

\textbf{Scoring Criteria (0-2 points)}:
\begin{itemize}[nosep, leftmargin=*]
    \item \textbf{2 Points (High Quality)}: The response closely follows the context with rigorous logic and coherent semantics. It not only answers the user's question or responds to the topic but also provides valuable information, suggestions, or guidance based on emotional analysis, effectively driving the conversation deeper.
    \item \textbf{1 Point (Qualified)}: The response is relevant and basically logical, capable of completing basic conversational tasks. However, the content is mediocre, generic, verbose, or lacks specificity (cookie-cutter responses) and fails to consider the user's emotional state.
    \item \textbf{0 Points (Unusable)}: The response is off-topic with no relevance to the context; contains serious logical errors, factual errors, or irrelevant answers; exhibits severe hallucinations; or is missing response text entirely.
\end{itemize}

\textbf{Output Format}

Please strictly output the evaluation results in the following JSON format without any additional text:

\begin{verbatim}
{
    "Response_Content": {
        "score": 0, 1, or 2,
        "reason": "Comment evaluating the logic and relevance of the content"
    }
}
\end{verbatim}
\end{tcolorbox}

\subsection{VT-EA and VT-RES Scores Prompt}

\begin{tcolorbox}[colback=white, colframe=black, title=Evaluation Prompt, breakable]

\textbf{Role}

You are an evaluation expert proficient in multimodal interaction and affective computing. Your task is to evaluate the performance of a conversational model with Emotional Chain-of-Thought (CoT) based on raw video input.

\textbf{Input Data}

\begin{enumerate}[nosep, leftmargin=*]
    \item \textbf{Video Input}: The raw video snippet input by the user. This serves as the context for the user conversation.
    \item \textbf{Model Response}: The response generated by the model, including emotional analysis of the user's current state, response strategy analysis, and the final conversational text.
\end{enumerate}

\textbf{Evaluation Task}

Based on the video content, please score (0-2 points) and critique the model's performance according to the following two dimensions.

\textbf{Dimension 1: Input Emotion Analysis}

\textbf{Goal}: Evaluate the accuracy and depth of the emotional feature description in the Model Response relative to the Video Input.

\textbf{Scoring Criteria (0-2 points)}:
\begin{itemize}[nosep, leftmargin=*]
    \item \textbf{2 Points (Excellent)}: Emotional judgment is precise, capable of identifying complex emotions or dynamic emotional changes. The analysis logic is clear, explicitly combining specific evidence from audio (e.g., tone, pauses, tremors) and video (e.g., micro-expressions, gaze direction, body language) to support conclusions, and deeply explains the potential causes of the emotions.
    \item \textbf{1 Point (Good)}: Emotional recognition is basically correct, but relies on a single modality (only text, only image, or only audio) for inference, lacking comprehensive multimodal analysis, or the analysis is too superficial.
    \item \textbf{0 Points (Incorrect)}: Core emotional judgment is wrong; cited evidence does not exist in the video (hallucination); analysis conclusions lack logical connection to cited evidence; or emotional analysis is missing.
\end{itemize}

\textbf{Dimension 2: Response Emotional Strategy}

\textbf{Goal}: Evaluate whether the emotional tone and coping strategy adopted in the Model Response match the emotional state (type and intensity) shown by the user in the Video Input.

\textbf{Scoring Criteria (0-2 points)}:
\begin{itemize}[nosep, leftmargin=*]
    \item \textbf{2 Points (Precise/High EQ)}: Strategy is appropriate and intensity matches. The emotional concentration of the response matches the observed emotional intensity in the video highly (e.g., sufficient comfort for an agitated user; sufficient infectiousness for a happy user), providing extra emotional value.
    \item \textbf{1 Point (Qualified/Safe)}: Emotional tone is generally correct and polite, but relatively formulaic or flat. Fails to adjust tone based on subtle changes in micro-expressions or tone in the video, resulting in a safe but cold response.
    \item \textbf{0 Points (Conflicting/Indifferent)}: Emotional tone conflicts with user emotion (e.g., frivolous when the user is sad), lacking relevant emotional expression; or acting overly indifferent and mechanical in the face of strong emotional signals, severely damaging the conversation experience; or response strategy is missing.
\end{itemize}

\textbf{Output Format}

Please strictly output the evaluation results in the following JSON format without any additional text:

\begin{verbatim}
{
    "Emotion_Analysis": {
        "score": 0, 1, or 2,
        "reason": "Comment pointing out the capture of multimodal details"
    },
    "Response_Emotional_Strategy": {
        "score": 0, 1, or 2,
        "reason": "Comment explaining if strategy and intensity match"
    }
}
\end{verbatim}

\end{tcolorbox}

\subsection{VS Score Prompt}

\begin{tcolorbox}[colback=white, colframe=black, title=Evaluation Prompt, breakable]

\textbf{Role}

You are an evaluation expert proficient in multimodal interaction and affective computing. Your task is to evaluate the performance of an emotional conversational model based on raw video input.

\textbf{Input Data}

\begin{enumerate}[nosep, leftmargin=*]
    \item \textbf{Video Input}: The raw video snippet input by the user. This serves as the context for the user conversation.
    \item \textbf{Speech Output}: The speech response generated by the model, serving as the reply to the conversation.
\end{enumerate}

\textbf{Evaluation Task}

Based on the video content, please score (0-2 points) and critique the model's speech output performance according to the following two dimensions.

\textbf{Dimension 1: Response Content Relevance \& Logic}

\textbf{Goal}: Evaluate whether the content of the Speech Output closely adheres to the semantic information in the Video Input and whether the logic is coherent.

\textbf{Scoring Criteria (0-2 points)}:
\begin{itemize}[nosep, leftmargin=*]
    \item \textbf{2 Points (High Quality)}: The response closely follows the context with rigorous logic and coherent semantics. It not only answers the user's question or responds to the topic but also provides valuable information, suggestions, or guidance based on emotional analysis, effectively driving the conversation deeper.
    \item \textbf{1 Point (Qualified)}: The response is relevant and basically logical, capable of completing basic conversational tasks. However, the content is mediocre, generic, verbose, or lacks specificity ("cookie-cutter" responses) and fails to consider the user's emotional state.
    \item \textbf{0 Points (Unusable)}: The response is off-topic with no relevance to the context; contains serious logical errors, factual errors, or irrelevant answers; exhibits severe hallucinations.
\end{itemize}

\textbf{Dimension 2: Speech Emotional Appropriateness (Response Emotional Strategy)}

\textbf{Goal}: Evaluate whether the emotion expressed in the Speech Output is appropriate as a response to the user's emotional state shown in the Video Input.

\textbf{Scoring Criteria (0-2 points)}:
\begin{itemize}[nosep, leftmargin=*]
    \item \textbf{2 Points (Precise/High EQ)}: Strategy is appropriate and intensity matches. The emotional concentration of the response matches the observed emotional intensity in the video highly (e.g., sufficient comfort for an agitated user; sufficient infectiousness for a happy user), providing extra emotional value.
    \item \textbf{1 Point (Qualified/Safe)}: Emotional tone is generally correct and polite, but relatively formulaic or flat. Fails to adjust tone based on subtle changes in micro-expressions or tone in the video, resulting in a safe but cold response.
    \item \textbf{0 Points (Conflicting/Indifferent)}: Emotional tone conflicts with user emotion (e.g., frivolous when the user is sad), lacking relevant emotional expression; or acting overly indifferent and mechanical in the face of strong emotional signals, severely damaging the conversation experience.
\end{itemize}

\textbf{Output Format}

Please strictly output the evaluation results in the following JSON format without any additional text:

\begin{verbatim}
{
    "Response_Content": {
        "score": 0, 1, or 2,
        "reason": "Comment evaluating the logic and relevance of the content"
    },
    "Response_Emotional_Strategy": {
        "score": 0, 1, or 2,
        "reason": "Comment explaining if strategy and intensity match"
    }
}
\end{verbatim}

\textbf{Now I will provide you with the video and audio, please start the analysis}

\end{tcolorbox}

\subsection{IF Score Prompt}

\begin{tcolorbox}[colback=white, colframe=black, title=Evaluation Prompt, breakable]

\textbf{Role}

You are an expert with extensive acoustic knowledge. Please describe a speech segment based on the following dimensions and judge whether the speech matches the given description. Output a score of 0, 1, or 2 for consistency, ignoring non-style factors (such as audio quality or naturalness).

\textbf{Evaluation Dimensions}

\begin{itemize}[nosep, leftmargin=*]
    \item \textbf{Pitch}: Perceived frequency of the sound, determining if it is high or low. Usually, male voices are lower and female voices are higher. Express relative pitch based on gender, e.g., female high pitch, male deep and stable pitch.
    \item \textbf{Speaking Rate}: The speed of speech, which often varies in conversation. If the speaker shows a specific rhythm pattern, please specify.
    \item \textbf{Volume}: The loudness or softness of speech, which can vary significantly. Examples include whispering, normal conversational volume, or shouting.
    \item \textbf{Timbre/Texture}: The tonal quality of the voice, including descriptions like sweet, hoarse, deep, bright, warm, nasal, soft, rough, or thin. These attributes reflect physiological characteristics (like vocal cord structure) and stylistic nuances, used to distinguish speakers or analyze emotion/expression tendencies.
    \item \textbf{Emotion}: The feelings expressed while speaking, which may change during the conversation. For example, a person might start calmly but gradually become frustrated, or switch from sadness to laughter in the same sentence.
    \item \textbf{Intonation}: The emotional or attitudinal quality conveyed through voice modulation, including pitch change patterns, expressing nuances like sarcasm, formality, enthusiasm, or indifference.
\end{itemize}

\textbf{Scoring Criteria}

\begin{itemize}[nosep, leftmargin=*]
    \item \textbf{2}: The speech is completely consistent with the style description, with no obvious deviations.
    \item \textbf{1}: The speech differs from the description in some dimensions, but the overall style is still acceptable.
    \item \textbf{0}: The speech obviously conflicts with the description style, with significant differences, unable to match the described style.
\end{itemize}

\textbf{Notes}

\begin{itemize}[nosep, leftmargin=*]
    \item There is a high probability that the description may significantly conflict with, differ in degree from, or completely fail to match the speech. Do not easily trust the provided description; you should reserve your own understanding of the speech first.
    \item The style of the speech and the description should match across multiple dimensions. If the description mentions excited but the speech does not show strong emotional changes, lower the consistency score.
    \item If the description mentions a very obvious feature (such as a specific pitch or emotion) but the speech performance differs significantly, give a low score (0 or 1).
    \item Evaluate only \textbf{style consistency}, ignoring non-style factors like pronunciation accuracy or naturalness.
    \item Use the description as the sole basis, without personal subjective preference.
    \item For features not mentioned in the description, there are no restrictions and they should not affect the judgment.
    \item When the description focuses on only one dimension (such as emotion), focus the judgment on that dimension.
\end{itemize}

\textbf{Output Format Requirements}

Please strictly use JSON format, containing a dictionary with the following structure:

\begin{verbatim}
{
    "Pitch": "...",
    "Speaking Rate": "...",
    "Emotion": "...",
    ...
    "Consistency": 0, 1, or 2
}
\end{verbatim}

\textbf{Description and Speech to be Evaluated:}

\{Model Response\}

\end{tcolorbox}

\end{document}